# Light-induced size changes in BiFeO$_3$ crystals

B. Kundys[1]*, M. Viret[1], D. Colson[1] and D. O. Kundys[2]

**Multifunctional oxides are promising materials because of their fundamental physical properties as well as their potential in applications[1]. Among these materials, multiferroics exhibiting ferroelectricity and magnetism are good candidates for spin electronic applications using the magnetoelectric effect, which couples magnetism and ferroelectricity. Furthermore, because ferroelectrics are insulators with a reasonable bandgap, photons can efficiently interact with electrons leading to photoconduction or photovoltaic effects[2,3]. However, until now, coupling of light with mechanical degrees of freedom has been elusive, although ferroelasticity is a well-known property of these materials. Here, we report on the observation, for the first time, of a substantial visible-light-induced change in the dimensions of BiFeO$_3$ crystals at room temperature. The relative light-induced photostrictive effect is of the order of 10$^{-5}$ with response times below 0.1 s. It depends on the polarization of incident light as well as applied magnetic fields. This opens the perspective of combining mechanical, magnetic, electric and optical functionalities in future generations of remote switchable devices.**

Vigorous research efforts on magnetoelectric effects in recent years [4-6] have stimulated the exploration of these properties in many materials[7] and revived the investigation of known compounds, of which BiFeO$_3$ stands out as the most important example[8]. Among the constantly growing, but still limited, number of materials combining ferroelectricity and (anti/ferro/ferri)magnetism, BiFeO$_3$ (BFO) occupies a unique place, as it is still the only compound with ferroelectricity and antiferromagnetism coexisting at room temperature. Although BFO has been known for a long time, it remains a subject of intensive investigations aiming to improve and understand its properties in bulk and thin-film forms[9]. As far as applications are concerned, BFO has a great potential in spintronics using the magnetoelectric effect to control a magnetization with an electric field. This property is indeed of great interest to write magnetic memories (for example, in magnetic random-access memories), efficiently and without significant heat production. As present `classical' spintronic components are based on metallic materials for which interaction with photons is very limited, functionalities with light have only recently been proposed in hybrid semiconductor systems[10]. These require materials that have a bandgap in which photons can generate charge carriers, for example, semiconductors and Schottky barriers. Interestingly, ferroelectric materials also offer this opportunity along with ferroelasticity, which couples mechanical strain to ferroelectric polarization. Hence, the great potential of magnetoelectric compounds for multifunctional applications, including magnetoelectric memory storage[11] and electric-field control of magnetic sensors, may successfully be extended to include other interesting effects such as magnetic- field dependent second-harmonic generation[12] or the photovoltaic effect that was recently discovered in BFO (refs 2,3). Here we report that ferroelectric single-domain BFO single crystal changes dimensions when subjected to light irradiation. Although this phenomenon, known as photostriction, has already been observed in a few materials (ferroelectric compounds[13-15], semiconductors[16,17] and polymers[18-20]), it has never been seen in (anti/ferro/ferri)magnetic compounds nor in magnetoelectric ones. Despite the potential technological interest in photostrictive materials, these have not been widely explored for photoelastic applications. Besides the scarcity of suitable materials, these also suffer from a too slow response time for implementing photoelastic devices. One can also wonder whether an intrinsic coupling with an additional external parameter, such as magnetic and/or electric fields, could influence the photostrictive response time or the magnitude of the effect. It is therefore interesting to investigate the possible existence of photostriction in a magnetoelectric compound. This is the goal of the present work, where we demonstrate that BiFeO$_3$ is indeed photostrictive at room temperature with a magnetic-field dependence of the photoelastic effect. This opens a new route for multifunctionality where strain, magnetization and polarization can potentially be changed simultaneously by light and applied magnetic and electric fields.

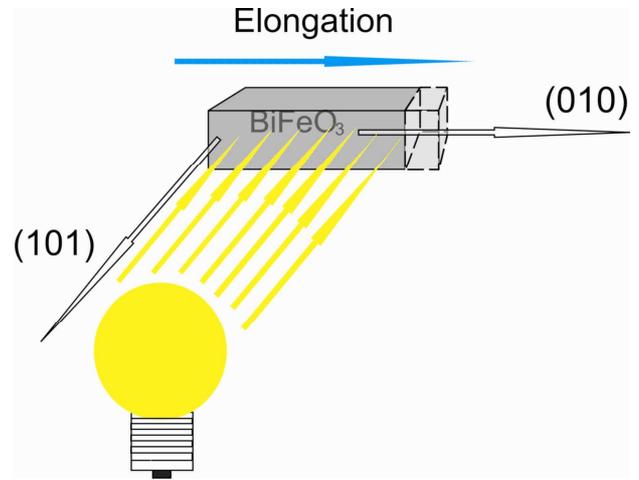

**Figure 1 Schematic of the photoelastic stress measurements**. Light irradiation and BiFeO$_3$ are shown with respect to crystalline directions.

A 1.6-mm-long, 0.3-mm-wide and 0.1-mm-thick crystal of BiFeO$_3$ containing a single ferroelectric domain was selected for this study. The sample was observed using polarized light with an optical microscope and was determined to be in a single domain state with its spontaneous ferroelectric polarization along the (111) direction (in pseudocubic lattice description). A HeNe laser

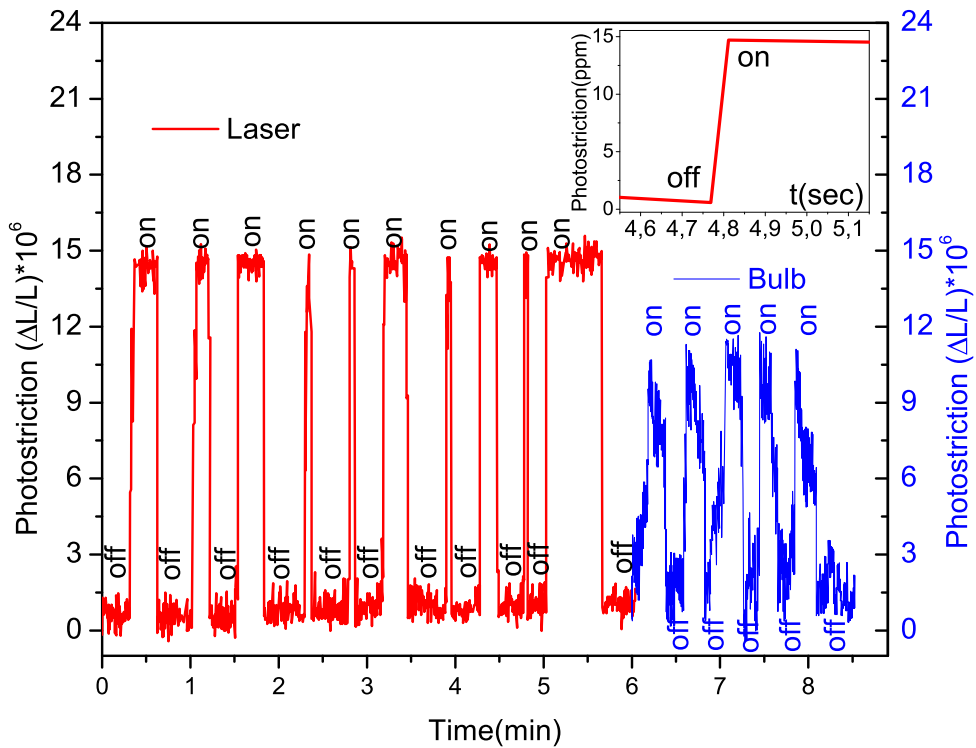

**Figure 2 Time dependence of photostriction in BiFeO₃. Different illumination times along the (010) crystalline direction were applied.** The measurements were carried out in darkness with HeNe laser light (15mW, 632.8 nm) in red, and using an ordinary continuous-spectrum, Philips white light bulb (230V, 100W) in blue. The inset is a zoom demonstrating the fast response time below 0.1 s.

(15mW, 632.8 nm) and an ordinary Philips white light bulb (230 V, 100 W) were used for illumination in darkness. The surface of the sample (0,48mm$^2$) was covered by light uniformly and photostriction was measured using a capacitance dilatometer. This home-made instrument, similar to that described in ref. 21, uses a capacitance bridge (at a 1 V oscillation voltage amplitude) to precisely measure any variation of distance between a fixed and a mobile electrode attached to one end of the sample. In our experiment, length changes of the BFO crystal along its (010) direction (the longest for this sample) were monitored while light was irradiated on and off along the (101) crystalline axis (see Fig. 1). A magnetic field could also be applied normal to the light propagation, along the (010) direction. The sample was not contacted with electrodes as this would reduce the measurement sensitivity. As can be seen in Fig. 2 (red line), the result of periodic sample illumination with the HeNe laser leads to a clear photoelastic effect, manifesting itself in a repeatable change of the sample length in the (010) direction. The stress produced by light is tensile with a relative change in the sample dimension of $14*10^{-6}$. The measured response time is below 0.1 s, which is much faster than that of some polymer materials[20]. Interestingly, this rules out a potential influence of temperature-driven expansion, as thermal effects vary much slower (it takes about 60s for similar BFO crystals to stabilize in temperature under red light illumination in ref. 2). The effect can even be seen under sample illumination using an ordinary, continuous-spectrum, white light bulb (230V, 100 W) (Fig. 2, blue line). Photo-induced strain can have a different origin in different materials, but it has been understood in ferroelectrics as a superposition of photovoltaic and converse piezoelectric effects. Indeed, as light impinges on a ferroelectric material, electron-hole pairs are generated. The internal electric field, originating from ferroelectric polarization, tends to separate these charges and produces a voltage. This is the photovoltaic effect and it has recently been evidenced in BFO (ref. 2). The electric field generated by this process is in the direction of the local polarization. This field is in turn responsible for creating a strain in the material owing to piezoelectric effects. In BFO, we find that photostriction depends on the polarization of incident light, in agreement with what has been reported for the photovoltaic effect[2]. Figure 3 shows a sinusoidal angular dependence of the sample's deformation with the light polarization rotation. The maximum elongation is obtained for a light polarization parallel to the in-plane component of the sample's polarization. Multiferroics also possess a magnetic degree of freedom that could potentially couple to either electron-hole pair generation or ferroelasticity.

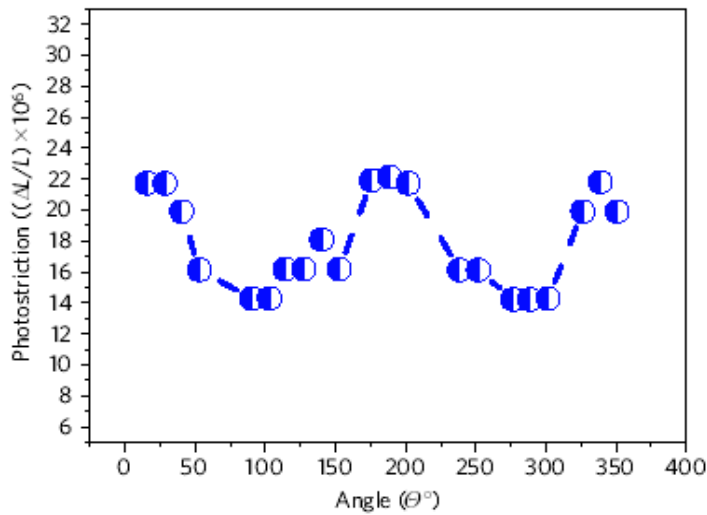

**Figure 3 Variation of photostriction with light polarization rotation.** The measurements were done under constant illumination using linearly polarized light from a HeNe laser. The effect is maximum when light polarization is along the in-plane component of the ferroelectric polarization and minimum in the perpendicular configuration.

Therefore, we have measured the influence of a magnetic field on photostriction (Fig. 4, inset) and we have found that the magnitude of the photoelastic coupling varies linearly with magnetic field (Fig. 4) to reach a 30% decrease in 1.5 T. Such a variation is hard to understand because a field of this magnitude only weakly affects an antiferromagnet such as $BiFeO_3$. Indeed, we have measured longitudinal magnetostriction along the (010) direction, and found it to be at least an order of magnitude below the photostrictive distortion. Moreover, we have also checked that a 1.5 T magnetic field does not noticeably affect the value of electrical polarization (our P(E) loops are unchanged when a magnetic field is applied).

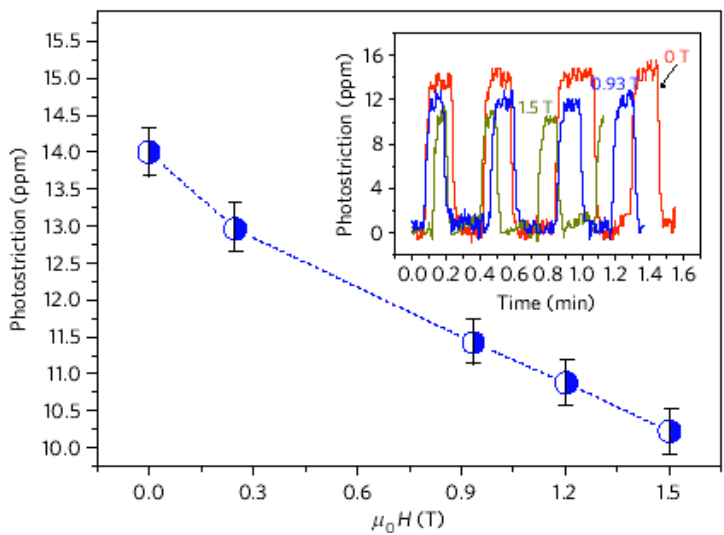

**Figure 4 Photostriction as a function of magnetic field.** The inset shows the distortion with periodic illumination (the period is not relevant here) using a HeNe laser (15mW, 632.8 nm) under three different magnetic fields. This property is most likely due to charge (electrons, holes) deviation under magnetic field.

In attempt to understand this magnetic field dependent photoelastic effect one can imagine that magnetic field deviates light induced electric charges (electrons, holes) that reduce their contribution to the polarization of the sample. Since electrostrictive and photovotaic effects may depend on the magnitude of polarization the phtotoelastic effect may become smaller. Nevertheless, there is a clear need for some deeper theoretical insight to understand the observed magnetic-field effect on photostriction. Importantly, we would like here to underline that the reported photostriction effect is most unlikely to originate from a light-induced temperature increase, although we were not able to directly measure the sample temperature. Indeed, besides the fast observed response times, the photostriction dependencies with both incident light polarization and magnetic field are inconsistent with a concomitant significant variation of absorbed light. We report here on photostrictive properties in a multiferroic compound. We find that $BiFeO_3$ expands under illumination with visible light. A relative elongation of $10^{-5}$ is observed when a 633-nm-wavelength laser light is shone onto a single crystal at a power of $\approx 70 kWm^{-2}$ ($\approx 470 Wm^{-2}$ for the bulb). This property is most likely due to a combination of photovoltaic and electrostrictive effects. The magnitude of photostriction also depends on the direction of light polarization and applied magnetic fields. Thus, $BiFeO_3$, the only robust multiferroic at room temperature, is definitely a most interesting material where polarization, magnetism, light and mechanical distortion all interact. This opens the way for potential applications in several types of wireless device and sensor, including light-controlled elastic micromotors, microactuators and even perhaps, other interesting optomechanical systems [22,23].

**Acknowledgements**

We acknowledge support from the French contracts: MELOIC (ANR-08-P196-36) of the `Agence Nationale de la Recherche' and BALISPIN (FF2008) from the `CNano Ile de France'. Author contributions The idea to measure photostriction in BiFeO3 belongs to B.K. Experiments were carried out by B.K. under supervision and participation of M.V. B.K. and M.V. wrote the Letter. Samples were prepared by D.C. D.O.K. helped with discussion and manuscript writing. Correspondence and requests for materials should be addressed to *kundys A gmail.com*